\documentclass[prd,onecolumn,preprintnumbers,nofootinbib]{revtex4}
\usepackage{graphicx}
\usepackage{pdfpages}
\usepackage{color}
\usepackage{float}

\usepackage{amsfonts,amsmath,amssymb}
\usepackage[plainpages=false, colorlinks=true, anchorcolor=blue, linkcolor=blue, citecolor=blue, bookmarks=false]{hyperref}
\usepackage{natbib}
\usepackage{enumitem}
\usepackage{lipsum}
\usepackage{multirow}
\usepackage{array}
\pdfoutput=1
\begin{document}
\newcommand{\bthis}[1]{\textcolor{black}{#1}}
\newcommand{\apjl}{Astrophys. J. Lett.}
\newcommand{\apjs}{Astrophys. J. Suppl. Ser.}
\newcommand{\aap}{Astron. \& Astrophys.}
\newcommand{\nar}{New  Astronomy Reviews}

\newcommand{\aj}{Astron. J.}
\newcommand{\araa}{Ann. Rev. Astron. Astrophys. } 
\newcommand{\mnras}{Mon. Not. R. Astron. Soc.}
\newcommand{\ssr}{Space Science Revs.}
\newcommand{\apss}{Astrophysics \& Space Sciences}
\newcommand{\jcap}{JCAP}
\newcommand{\pasj}{PASJ}
\newcommand{\pasp}{PASP}
\newcommand{\pasa}{Pub. Astro. Soc. Aust.}
\newcommand{\physrep}{Phys. Rep.}
\renewcommand{\arraystretch}{2.5}
\title{Calibration of luminosity correlations of gamma-ray bursts using quasars}
\author{Sarveshkumar \surname{Purohit}}\altaffiliation{sarveshpurohit84@gmail.com}

\author{Shantanu  \surname{Desai}}  
\altaffiliation{shntn05@gmail.com}

\affiliation{Department of Artificial Intelligence, Indian Institute of Technology, Hyderabad, Telangana-502284, India}

\affiliation{Department of Physics, Indian Institute of Technology, Hyderabad, Telangana-502284, India}

\begin{abstract}
In order to test the efficacy of   Gamma-ray Bursts (GRB)   as cosmological probes, we  characterize the scatter in the  correlations between six pairs of GRB observables, which have previously also been studied in ~\cite{Li21}.  However,  some of these observables depend on the luminosity distance, for which one needs to   assume an underlying cosmological model.
In order to circumvent this circularity problem, we use X-ray and UV fluxes of quasars as distance anchors to   calculate the luminosity distance in a model-independent manner, which in turn is used to calculate the GRB-related quantities. We find that all  six pairs of  regression relations show  high intrinsic scatter for both the low and high redshift sample. This implies that these GRB observables cannot be used as model-independent high precision cosmological probes.
\end{abstract}

\maketitle

\section{Introduction}
Gamma-ray bursts (GRBs) are single-shot explosions located at cosmological distances, which were first detected in 1960s  and have been observed over ten  decades in energies from  keV to over 10~TeV range~\cite{Kumar,WuGRBreview}.  They are located at cosmological distances, although a distinct time-dilation signature in the light curves is yet to be demonstrated~\cite{Koceski,Butler,Singh}. Because of their high energies and cosmological distances, they have also proved to be  very good probes of fundamental physics, such as testing  of Lorentz invariance violation and quantum gravity~\cite{AmelinoCamelia98,Abdo,Vasileiou13,Vasileiou15,Desai23}. GRBs are traditionally divided into two categories based on their durations, with  long (short) GRBs lasting more (less) than two seconds~\cite{Kouveliotou}.
Long GRBs are usually associated with core-collapse supernovae~\cite{Woosley} and short GRBs with neutron star mergers~\cite{Nakar}. There are however many exceptions to this conventional dichotomy~\cite{Gehrels,Ahumada,Troja22,Yang24} and many claims for additional GRB sub-classes have also been discussed in literature~\cite{Norris06,Kulkarni,Horvath,Bhave}. 

GRBs have  been proposed  as distance indicators or standard candles for cosmological purposes, over the past two decades due to putative  correlations between myriad   GRB observables in both the prompt and afterglow emission phase~\cite{DainottiAmati,Ito,Delvecchio,Luongo,Dainotti22,PradyumnaGRB}.  These correlations have often been used to estimate cosmological parameters~\cite{Moresco}. An uptodate review of all GRB  correlations in the prompt and afterglow phase along with cosmological applications of these relations can be found in \cite{Dainotti24}.

However, there is an inherent circularity problem in using the GRB observables as cosmological probes, since the calculation of  these observables is based on an underlying cosmological model. {However, in both the cases the correlation should be preserved. To get around this circularity problem, two methodologies have been used in literature. One way is to simultaneously constrain the GRB correlations along with the cosmological parameters~\cite{Khadka21,KhadkaRatra20}. Alternately,  a large number of ancillary probes have also  been used to obtain cosmology-independent estimates of distances corresponding to the GRB redshift such as Type Ia SN, cosmic chronometers, BAO $H(z)$ measurements, X-ray and UV luminosities of quasars, galaxy clusters etc~\cite{Govindaraj} (and references therein).

Here, we focus on correlations between six pairs of GRB observables: $\tau_{lag}-L$, $V-L$, $E_{p}-L$, $E_{p}-E_{\gamma}$, $\tau_{RT}-L$, $E_{p}-E_{iso}$, which were proposed in ~\cite{Wang11} (see also ~\cite{Basil}).
\footnote{We note that among the above correlations, the  $E_{p}-E_{\gamma}$ relation  is often referred to as as the Amati relation in literature~\cite{Amati06}. Also the correlation between $E_p$ and $L$ is also known as Yonetoku relation and has been widely studied in literature ~\cite{Yonetoku,Kodama}. All the other correlations were first considered in ~\cite{schaefer2007hubble}.}
The aforementioned  work simultaneously fitted  for cosmology along with  regression relations between the above parameters~\cite{Wang11}. Subsequently. these correlations were then studied  by obtaining model-independent distances to GRBs using the Pantheon compilation~\cite{Scolnic18} of Type Ia SN in ~\cite{Li21} (T21, hereafter).  Here, we carry out a variant of the analysis done in T21, by using the  X-ray and UV luminosities of quasars instead of Type Ia SN to probe the same correlations first considered in ~\cite{Wang11}. We note that quasars have previously being used to probe the efficacy of the Amati relation~\cite{Dai21} for the GRB datasets in ~\cite{Khadka21,Demianski17,Demianski2}.

The outline of this manuscript is as follows. We discuss the datasets used in this work in Sect.~\ref{sec:datasets}. Our analysis and results can be found in Sect.~\ref{sec:GPR}. We conclude in Sect.~\ref{sec:conclusions}.

\section{Datasets}
\label{sec:datasets}
We use the same GRB datasets as those considered in T21, which were originally obtained from  ~\cite{Wang11}.  This work considered a sample of 116 long GRBs with redshifts between 0.17 to 8.2. This sample consisted of GRBs observed by SWIFT until 2012  in conjunction with  69 additional GRBs from other detectors  obtained before launch of the SWIFT  satellite~\cite{schaefer2007hubble}. The dataset in ~\cite{schaefer2007hubble} consisted  of long GRBs with well measured and robust redshifts. For the additional SWIFT GRBs considered in ~\cite{Wang11}, any long GRB with more 100\% error in any observable was discarded.
The observables assembled for each GRB consists of bolometric peak flux  ($P_{bolo}$), bolometric fluence ($S_{bolo}$), beaming factor ($F_{beam}$), time lag between low and high energy photon light curves ($\tau_{lag}$), peak energy of the spectrum ($E_{p}$), minimum rise time of the peaks for which the light curve rise by half its peak flux ($\tau_{RT}$), and the variability of the light curve ($V$), along with their $1\sigma$ error bars. 
Among these  measurements, $\tau_{RT}$, $E_{p}$, $\tau_{lag}$, and $V$ were obtained directly from spectral analysis, whereas $P_{bolo}$ and $S_{bolo}$ can be obtained from the observed GRB spectrum as outlined in ~\cite{Wang11}.  In a future work, we shall update this dataset using the entire SWIFT catalog. In order to test for potential correlations,
one needs to calculate  the isotropic  peak luminosity ($L$), isotropic equivalent energy ($E_{iso}$), collimation-corrected energy ($E_{\gamma}$), which in turn depend on the luminosity distance ($D_L$),   for which one needs to posit  an underlying cosmological model. 
The relation between $L$  and $P_{bolo}$ is given by: 
\begin{equation}
L=4\pi D_L^2 P_{bolo},
\label{eq:L}
\end{equation}
$E_{iso}$ is related to $D_L$ using:
\begin{equation}
E_{iso}= 4\pi D_L^2 S_{bolo}(1+z)^{-1}
\label{eq:Eiso}
\end{equation}
Finally $E_{\gamma}$  is given by:
\begin{equation}
E_{\gamma}=4\pi D_L^2 S_{bolo} F_{beam} (1+z)^{-1},
\label{eq:Egamama}
\end{equation}
where $F_{beam}$ is the beaming factor, which was estimated using the empirical formula derived in ~\cite{Sari99}.

This dataset was used to study six different pairs of luminosity correlations: $\tau_{lag}-L$, $V-L$, $E_{p}-L$, $E_{p}-E_{\gamma}$, $\tau_{RT}-L$, $E_{p}-E_{iso}$, where some of the above variables were scaled according to redshift as explained in the next section.   To circumvent this circularity problem, a combined  fit to a linear regression between the above variables and the underlying cosmology model was done~\cite{Wang11}. The intrinsic scatter of the $V-L$ correlation was found to be very large, but the other variables had a tight correlation  with a negligible redshift evolution.  These same set of correlations were again considered  in T21 using a model-agnostic approach without assuming an underlying cosmological model. For this purpose, a model-independent estimate of $D_L$ at each GRB redshift was obtained using deep learning and Gaussian process  based regression using Type Ia supernova data, which overlap in redshifts with the GRBs. T21 also tested for a redshift evolution for the same  six regression relations (considered in ~\cite{Wang11}) by dividing the GRB dataset into a low redshift and high redshift sample. Among these,  only $E_{p}-E_{\gamma}$ was found to have no redshift evolution~\cite{Li21}.

\subsection{Quasar dataset}
In order to calculate $D_L$ corresponding to a given GRB redshift, instead of Type Ia SN,  we use quasars as distance anchors. Quasars consist of  active galactic nuclei, where the energy release occurs due to the accretion onto a supermassive black hole.  Quasars have been detected up to a redshift of $z=7$ and therefore span the same redshift range  as  GRBs. Quasars have been observed  throughout the electromagnetic spectrum~\cite{Mortlock}. A tight scaling relation  between the optical-UV flux at rest frame  wavelength of 2500 \AA  and X-ray flux  at rest frame energy of 2 keV  of a quasar has been asserted, which is independent of redshift~\cite{Lusso}.
The value of $D_L$ at a given redshift is obtained from the quasar X-ray and UV flux as follows~\cite{Lusso,Lusso20}:
\begin{equation}
\log D_L = \frac{[\log F_X  -\beta -\gamma (\log F_{UV} + 27.5)]}{2 (\gamma-1)} -0.5 \log (4\pi)+28.5,
 \end{equation}
where  $F_X$ and $F_{UV}$ are the flux densities (in erg/s/\rm{$cm^2$}/Hz) 
$\beta \approx 8.2$ and $\gamma=0.6$.  This relation was obtained assuming a Hubble constant value of $H_0=70$ km/sec/Mpc.
Lusso et al~\cite{Lusso20} constructed a clean sample of  2,421 optically selected quasars spanning the redshift range $0.006 \leq z \leq 7.52$ (with a dispersion in the $L_X-L_{UV}$ relation of 0.24 dex),  where the distance modulus ($\mu$) and associated errors were obtained using $D_L$ from the above equation as below: 
\begin{equation}
\mu = 5 \log(D_L) - 5 \log (10 \rm{pc})
\label{eq:DM}
\end{equation}
Therefore one can obtain the distance modulus for each quasar redshift from the quasar X-ray and UV fluxes. We however note that concerns have been raised that this quasar  dataset~\cite{Lusso20} may be not be  standardizable~\cite{Ratra1,Ratra2,Ratra3}, which is probably caused by the dust extinction in the $L_X-L_{UV}$ relation~\cite{Ratra4}. Nevertheless, since the quasar dataset in ~\cite{Lusso20} have been used to test for GRB correlations~\cite{Dai21}, we use this QSO dataset as distance anchors to directly compare  the scatter obtained in T21 using Type Ia supernovae. More detailed studies with a pure quasar sample will be deferred to a future work.

\section{Analysis and Results}
\label{sec:GPR}
The first step in the analysis involves obtaining a model-independent estimate of $D_L$ from the quasar dataset. As mentioned in the previous section, we start with the  $\mu$ and $z$ for 2,421 quasars.  With this data, we carry out a non-parametric reconstruction of $\mu$ at any redshift $z$ using Gaussian Process Regression (GPR). GPR is a generalization of a Gaussian distribution, characterized by a mean and a covariance function (usually called the kernel function)~\cite{seikel12}. 
For the GPR implementation,  we use the publicly available Python package {\tt sklearn}~\cite{sklearn} to reconstruct the distance modules ($\mu$) as a function of redshift ($z$) as follows:
The kernel we have used for the regression  is a sum of linear and constant kernels. A linear kernel with an exponent captures relations in the data, and a constant kernel is used to scale magnitude.
The GPR reconstruction of $\mu$ along with associated  $1\sigma$ error bars can be found in Fig.~\ref{fig1}. Once we have reconstructed $\mu$ at any $z$, we can estimate  $D_L$  by inverting Eq.~\ref{eq:DM}. The errors in $D_L$ can be obtained using standard error propagation.  We then obtain $L$ and  $E_{iso}$  using Eq.~\ref{eq:L} and Eq.~\ref{eq:Eiso}, respectively.
\begin{figure}
\centering
\includegraphics[width=12cm,height=12cm,keepaspectratio]{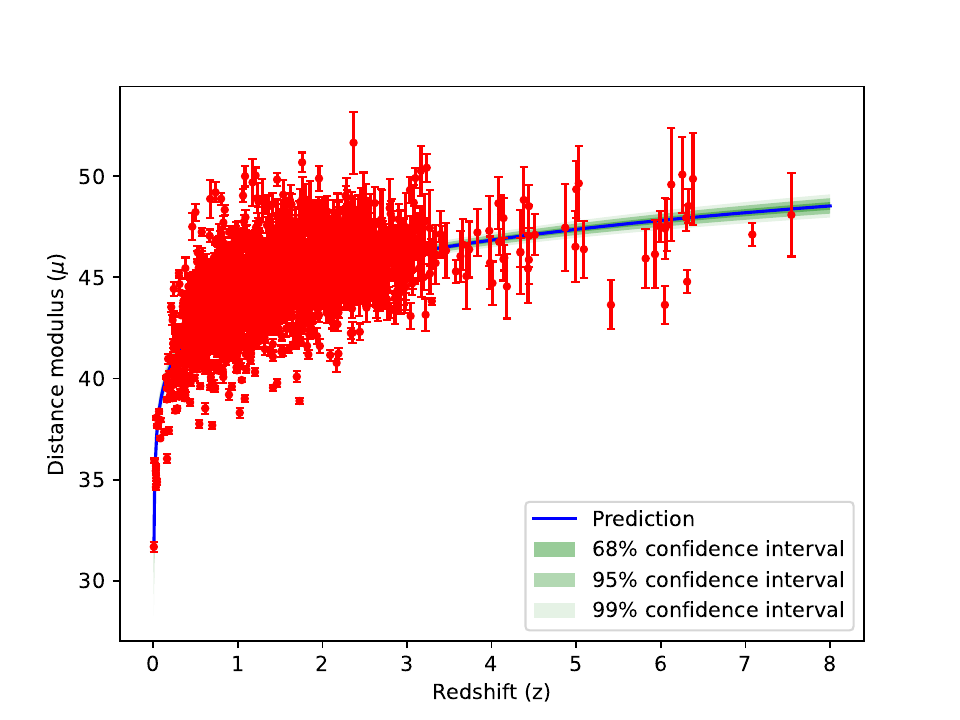}
\caption{Reconstruction of $\mu$ with GPR using X-ray and UV fluxes for 2,421 quasars.}
\label{fig1}
\end{figure}

We then carry out linear regression  between the six aforementioned GRB quantities in log space as follows:
\begin{align}
\log \frac{L}{\operatorname{erg} \mathrm{s}^{-1}} &= a_{1}+b_{1} \log \frac{\tau_{\operatorname{lag}, i}}{0.1 s}, \\
\log \frac{L}{\operatorname{erg~s}^{-1}} &= a_{2}+b_{2} \log \frac{V_{i}}{0.02}, \\
 \log \frac{L}{\operatorname{erg~s}^{-1}} &= a_{3}+b_{3} \log \frac{E_{p, i}}{300 \mathrm{keV}}\\
\log \frac{E_{\gamma}}{\text { erg }} &= a_{4}+b_{4} \log \frac{E_{p, i}}{300 \mathrm{keV}}, \\
\log \frac{L}{\text { erg s }} &= a_{5}+b_{5} \log \frac{\tau_{\mathrm{RT}, i}}{0.1 \mathrm{~s}}, \\
\log \frac{E_{\text {iso }}}{\text { erg }} &= a_{6}+b_{6} \log \frac{E_{p, i}}{300 \mathrm{keV}}
\end{align}
where  $\tau_{\mathrm{RT}, i}=  \tau_{\mathrm{RT}}/(1+z)$, $\tau_{\operatorname{lag}, i}= \tau_{\operatorname{lag}}/(1+z)$, $V_{i}=V(1+z)$ and $E_{p,i}=E_p(1+z)$.

To get the best-fit parameters  for each of the above equations, we apply the D’Agostini’s likelihood which incorporates errors in both the ordinate and abscissa~\cite{d2005fits}:
\begin{equation}
\mathcal{L}\left(\sigma_{\mathrm{int}}, a, b\right) \propto \prod_{i} \frac{1}{\sqrt{\sigma_{\mathrm{int}}^{2}+\sigma_{y i}^{2}+b^{2} \sigma_{x i}^{2}}} \times \exp \left[-\frac{\left(y_{i}-a-b x_{i}\right)^{2}}{2\left(\sigma_{\mathrm{int}}^{2}+\sigma_{y i}^{2}+b^{2} \sigma_{x i}^{2}\right)}\right],
\end{equation}
where $x$ and $y$ denote the abscissa and ordinate and $\sigma_{x i}$ and $\sigma_{y i}$ are the corresponding errors; 
$\sigma_{\mathrm{int}}$ denotes the intrinsic scatter in each regression relation. To get the best-fit values of each of the parameters, we do Bayesian regression and sample the posterior using the {\tt emcee} MCMC sampler~\cite{emcee}. We use uniform priors on both $a$ and $b$. Since we have obtained the $d_L$ using quasar UV and X-ray fluxes without an underlying cosmological model, we don't need to impose any priors on cosmological parameters.

In order to  study the evolution of the  aforementioned correlations   with redshift,  we  bifurcate  the GRB datset into two subsamples corresponding using the same redshift intervals as T21: the low-z sample ($z \leq 1.4$) which consists of 50 GRBs, and the high-z sample ($z > 1.4$) which consists of 66 GRBs.  We investigate the redshift dependence of luminosity correlations for these two subsamples. We also show the results for the full GRB sample.

The best-fit intervals at 68\%, 90\%, and 99\% credible intervals for all six regression relations can be found in Fig.~\ref{fig_contour}, for the full GRB sample and also after splitting the sample based on redshift. The scatter plots  between the six pairs of variables along with the best-fits are shown in Fig.~\ref{fig_xy}. We find that our results for the slope, intercept and intrinsic scatter for almost all the six scaling relations are consistent with that obtained in T21. The only exception is the $E_p-E_{\gamma}$ or Amati correlation, where we see an intrinsic scatter of 26-40\%, whereas the intrinsic scatter was found to be $<23\%$ in T21.  We find that the values for the slope and intercept are consistent or at most $1-2\sigma$ discrepant between the low redshift and high redshift samples indicating negligible evolution of the scaling relations.  Furthermore,  all the six regression relations show a high intrinsic scatter of greater than 30\%.  This includes the Amati relation, where we get a large scatter of 47\%, when we consider the full data sample, similar to our results of using galaxy clusters as distance anchors~\cite{Govindaraj}.  Therefore, the regression relations between these observables  using the quasar dataset in ~\cite{Lusso20} as distance anchors which has large dispersion cannot be used as model-independent probes of cosmological parameters. Another possible reason for the GRB dataset we have analyzed is not standardizable and one needs to check for that before trying to calibrate them~\cite{Ratra1,Ratra2,Ratra3}.


\begin{figure}[htbp]
\centering
\includegraphics[width=0.44\textwidth]{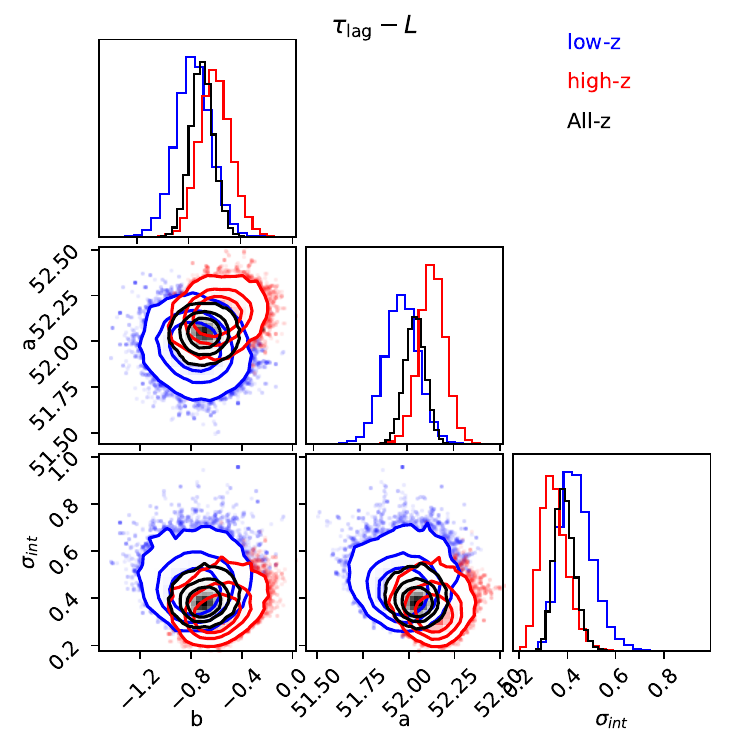}
\includegraphics[width=0.44\textwidth]{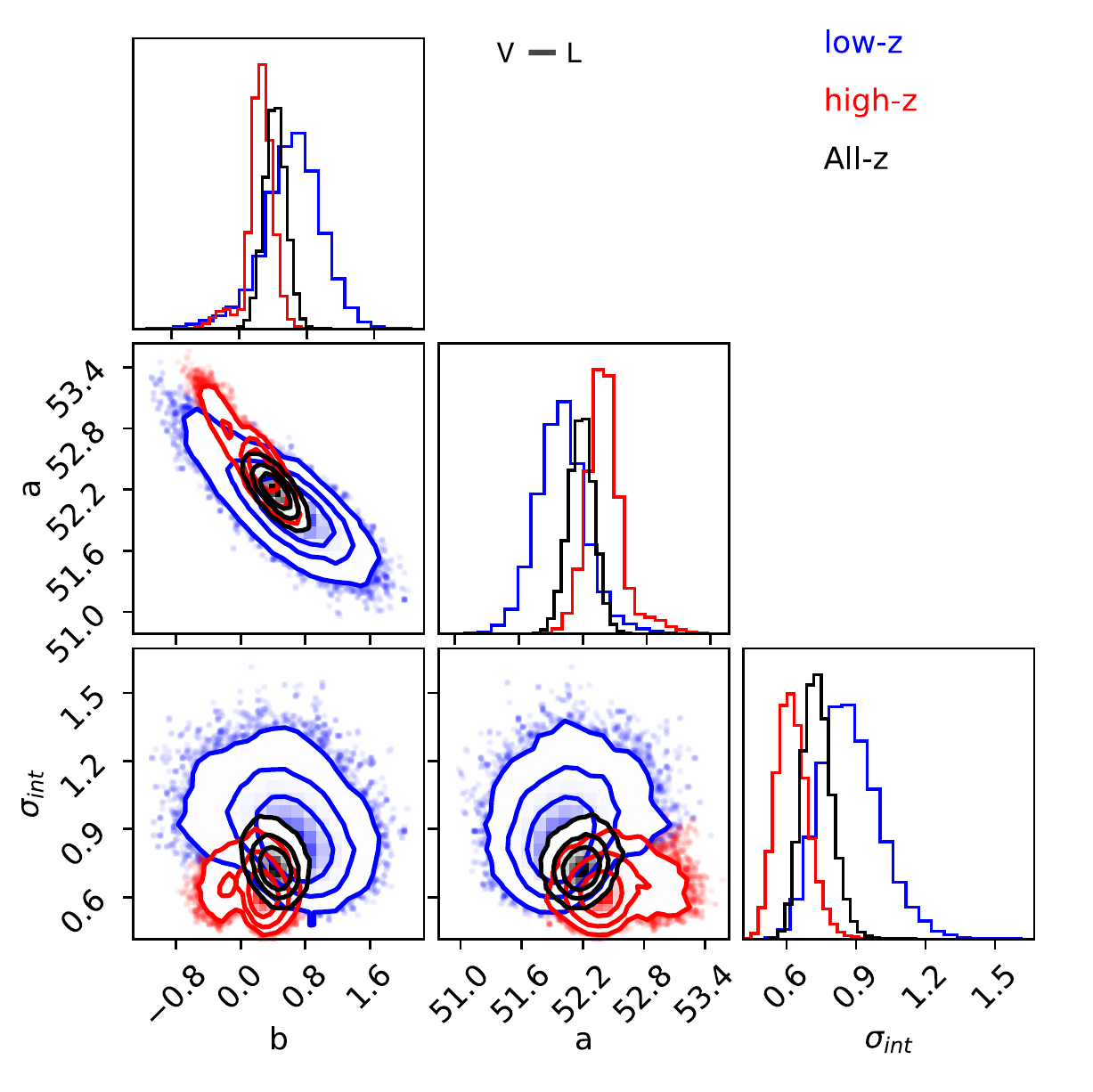}
\includegraphics[width=0.44\textwidth]{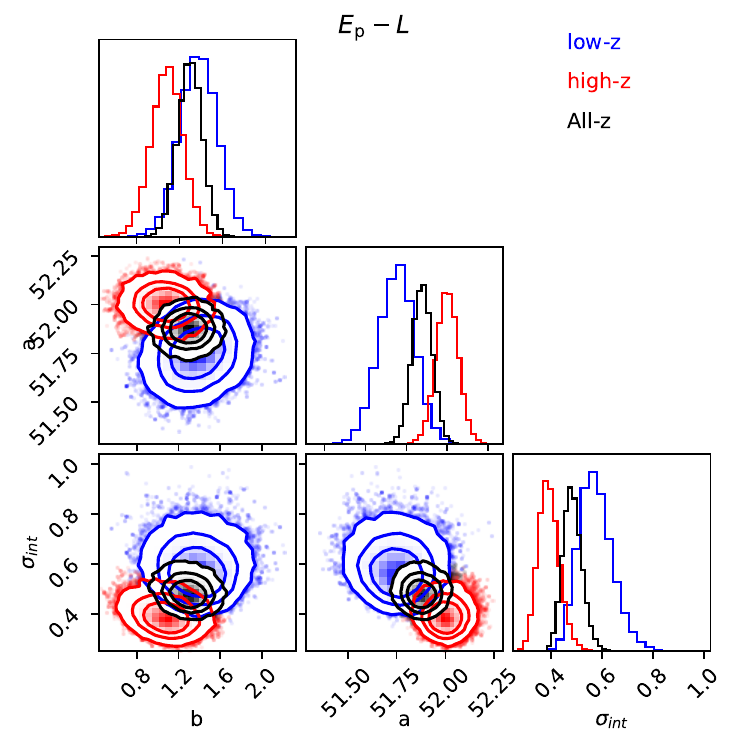}
\includegraphics[width=0.44\textwidth]{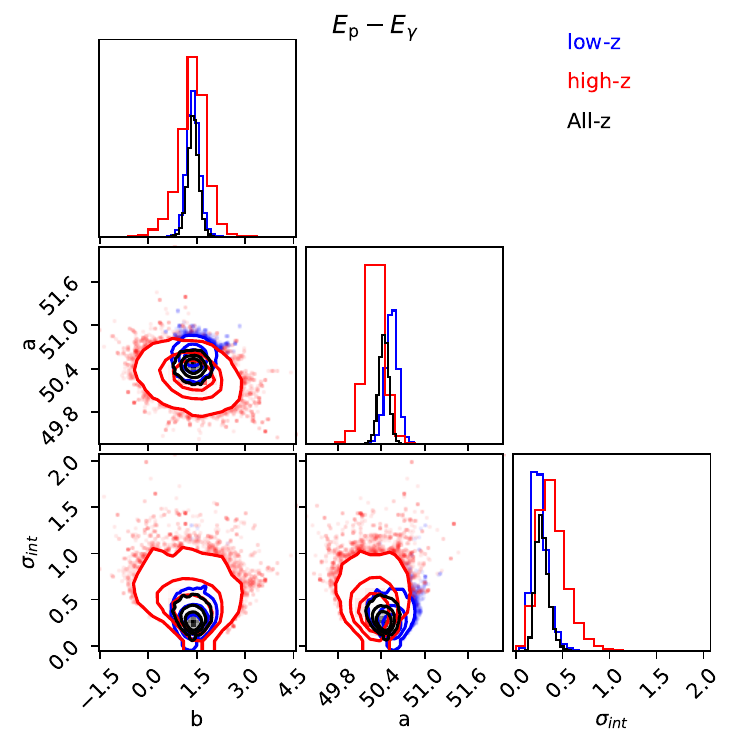}
\includegraphics[width=0.44\textwidth]{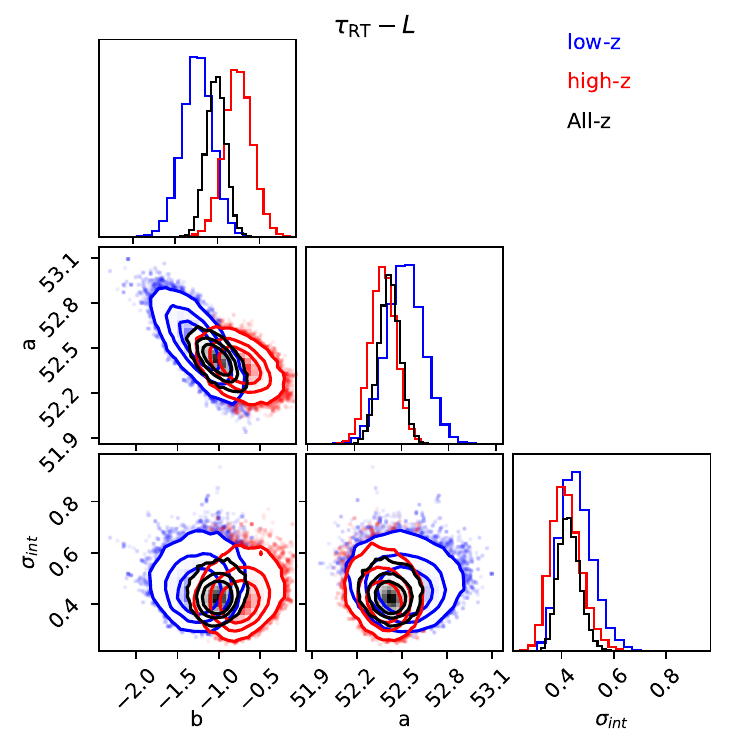}
\includegraphics[width=0.44\textwidth]{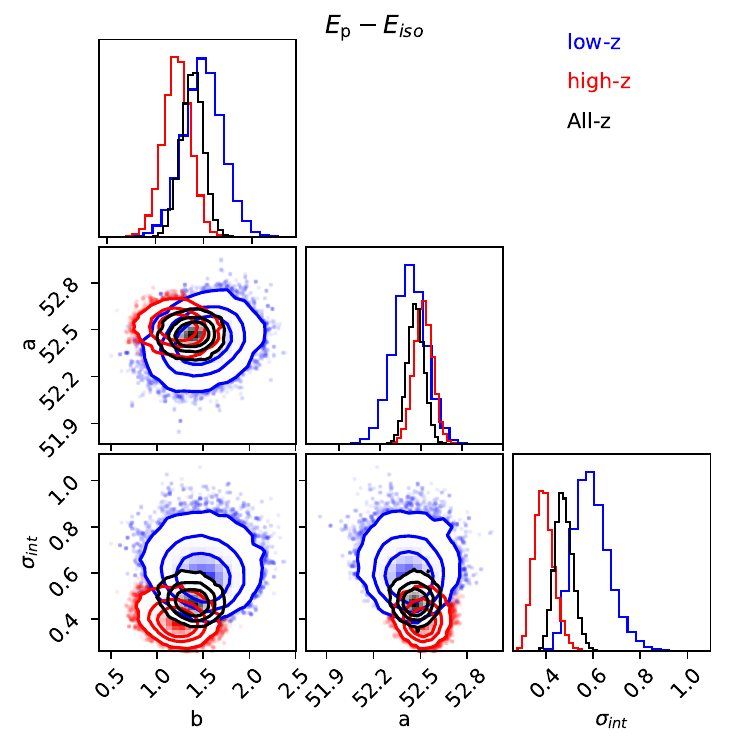}
\caption{\small{The 68\%,90\%, and 99\% credible intervals  along with the marginalized PDFs for the parameters for the regression relations between six pairs of GRB observables for low redshift ($z<1.4$)}, high redshift ($z>1.4$) as well as the full GRB dataset.}\label{fig_contour}
\end{figure}

\begin{figure}[htbp]
\centering
\includegraphics[width=0.48\textwidth]{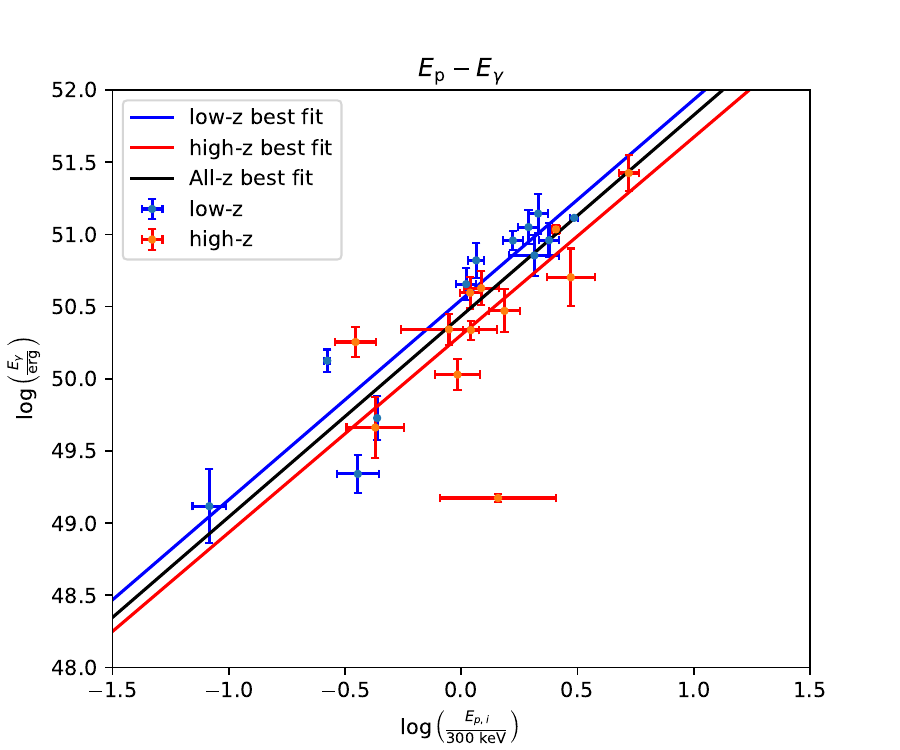}
\includegraphics[width=0.48\textwidth]{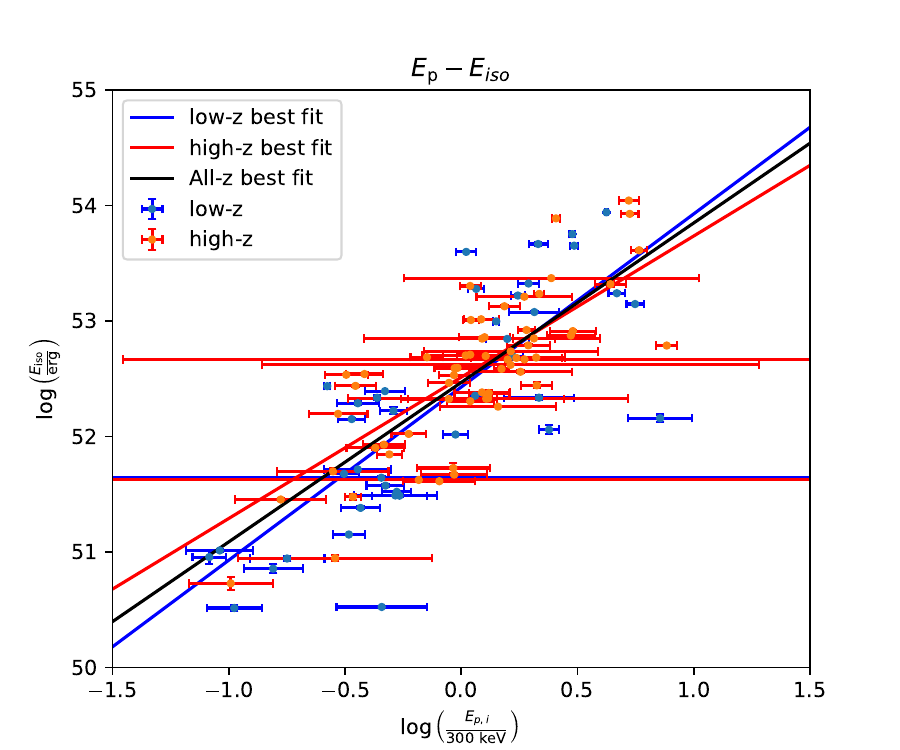}
\includegraphics[width=0.48\textwidth]{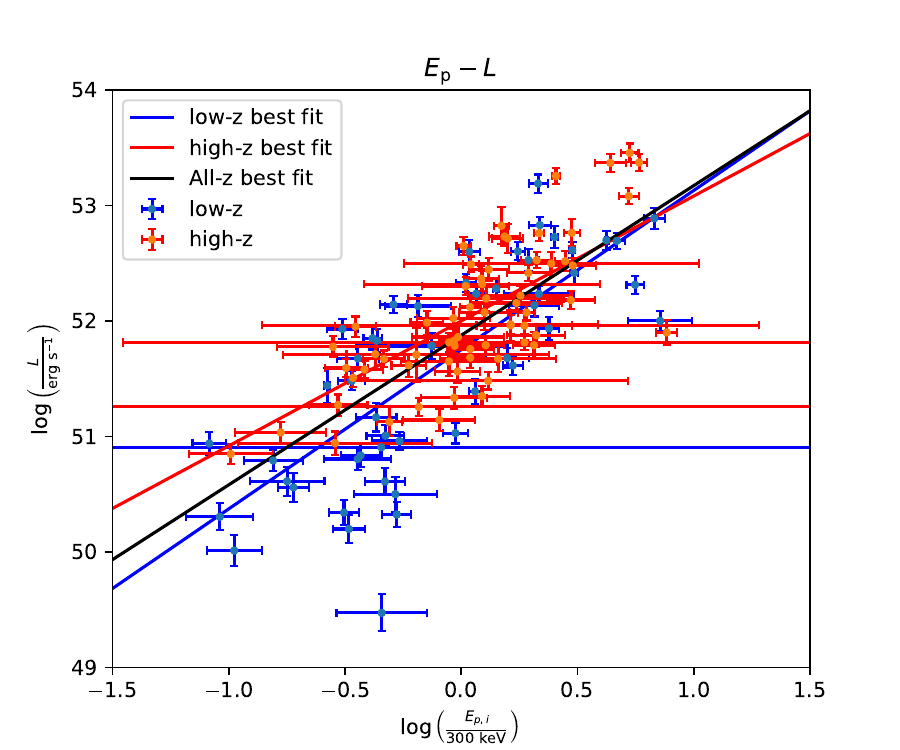}
\includegraphics[width=0.48\textwidth]{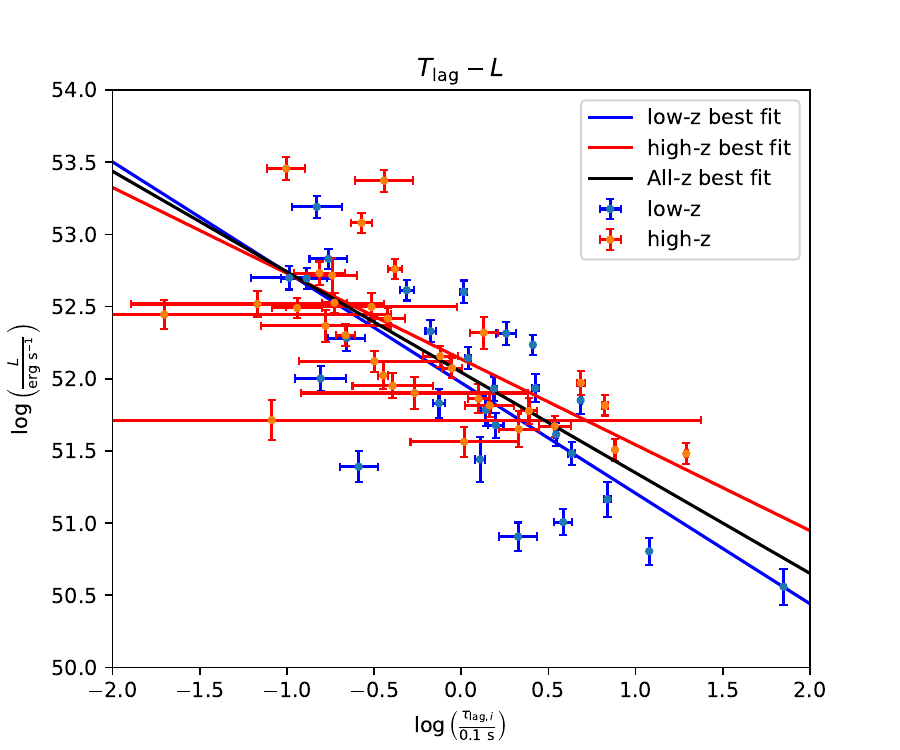}
\includegraphics[width=0.48\textwidth]{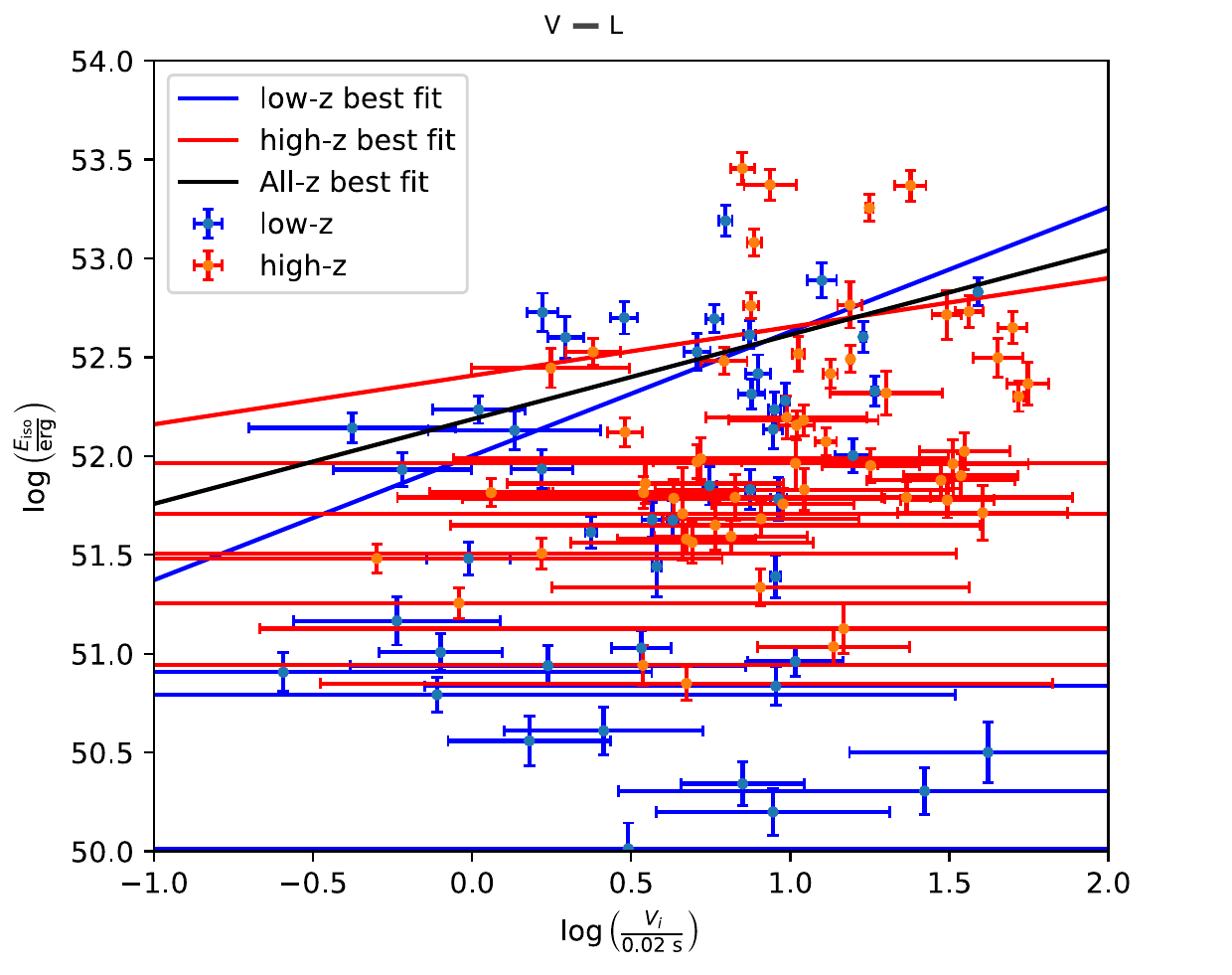}
\includegraphics[width=0.48\textwidth]{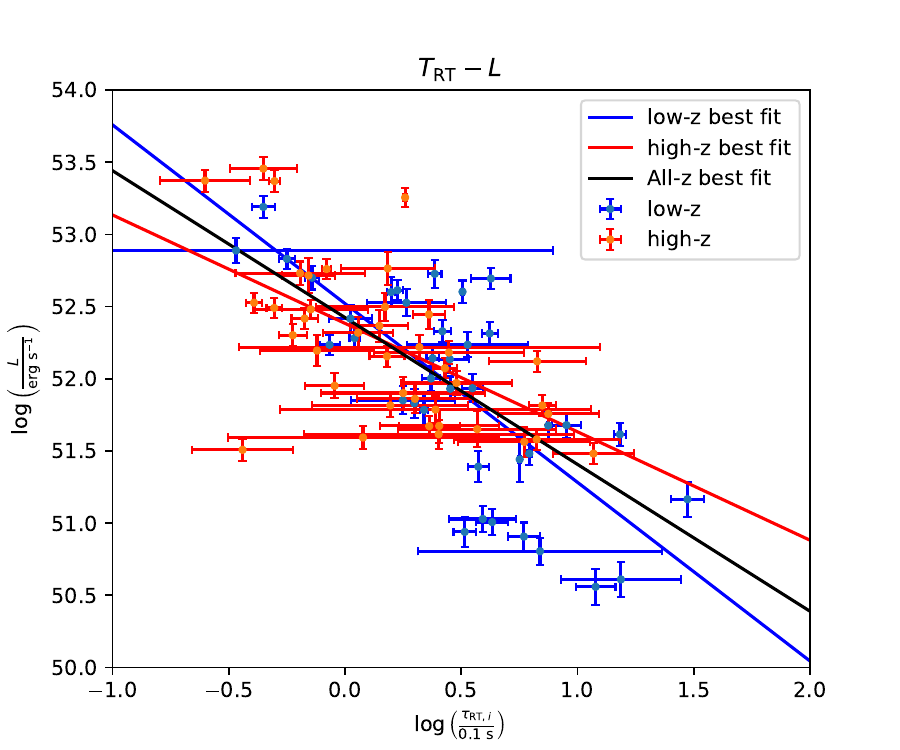}
\caption{\small{The luminosity correlations for low-$z$  and high-$z$  GRBs. The error bars denote the 1$\sigma$ uncertainties. The lines depict the best-fit results.}}\label{fig_xy}
\end{figure}

 \begin{table}[htbp]
\centering
\caption{\small{The best-fit  parameters of GRB luminosity correlations. $N$ is the number of GRBs in each subsample.}}\label{tab:parameters}
\arrayrulewidth=1.0pt
\renewcommand{\arraystretch}{1.3}
{\begin{tabular}{cccccc} 
\hline\hline 
Correlation &Sample & N  & $a$     & $b$   &$\sigma_{\rm int}$\\\hline

$\tau_{\rm lag}-L$ &low-z  &37 &51.97$\pm$0.09& 	$-0.77\pm0.14$ & 0.44$\pm$0.07\\
               &high-z &32 &52.14$\pm$0.07& 	$-0.59 \pm 0.12$& 	0.35$\pm$0.06\\
               &All-z &69 &52.04$\pm$0.06& 	$-0.69 \pm 0.09$& 	0.39$\pm$0.04\\
\hline
$V-L$ &low-z  &47 &52.00$\pm$0.24& 	0.63$\pm$0.36& 	0.88$\pm$0.13\\
      &high-z &57 &52.41$\pm$0.18& 	0.25$\pm$0.17& 	0.63$\pm$0.07\\
      &All-z &104 &52.19$\pm$0.12& 	0.43$\pm$0.13& 	0.73$\pm$0.06\\
\hline
$E_p-L$ 	&low-z  &50 &51.75$\pm$0.09& 	1.38$\pm$0.18& 	0.58$\pm$0.07\\
            &high-z &66 &51.99$\pm$0.06& 	1.08$\pm$0.15& 	0.39$\pm$0.04\\
            &All-z &116 &51.88$\pm$0.05& 	1.29$\pm$0.11& 	0.48$\pm$0.04\\
\hline
$E_p-E_{\gamma}$ &low-z  &12  &50.55$\pm$0.09& 	1.39$\pm$0.20& 	0.26$\pm$0.09\\
                 &high-z &12  &50.30$\pm$0.14& 	1.37$\pm$0.44&   0.39$\pm$0.16\\
                 &All-z &24   &50.43$\pm$0.07& 	1.39$\pm$0.17&   0.29$\pm$0.07\\
\hline
$\tau_{\rm RT}-L$ &low-z  &39  &52.52$\pm$0.12& 	-1.24$\pm$0.18& 0.46$\pm$0.06\\
              &high-z &40  &52.38$\pm$0.08& 	-0.75$\pm$0.17& 	0.42$\pm$0.06\\
              &All-z &79   &52.42$\pm$0.07& 	-1.02$\pm$0.11& 	0.43$\pm$0.041\\
\hline
$E_p-E_{\rm iso}$ &low-z  &40 &52.43$\pm$0.10& 	1.49$\pm$0.20& 0.59$\pm$0.08\\
              &high-z &61 &52.51$\pm$0.06& 	1.22$\pm$0.14& 	0.39$\pm$0.04\\
              &All-z &101 &52.47$\pm$0.05& 	1.38$\pm$0.12& 	0.48$\pm$0.04\\
\hline
\end{tabular}}
\end{table}

\section{Conclusions}
\label{sec:conclusions}
In a recent work, T21 studied  the empirical correlations among six pairs of GRB observables for 116 long GRBs (first considered in ~\cite{Wang11}) in order to test their efficacy as a cosmological probe. However, one needs an estimate of the  luminosity distance, which depends on an underlying cosmological model to calculate some of  these GRB observables. In order to get around this circularity problem, luminosity distances from Type Ia supernovae were used as distance anchors, and the corresponding distance at a given GRB redshift was obtained using Artificial Neural Network based regression.

In this work, we carry out the same exercise as T21, but use X-ray and UV fluxes of quasars instead of Type Ia supernovae in order to get the  luminosity distance 
at a given GRB redshift. The interpolation has been done using Gaussian process regression. Similar to T21, we test the correlations for both the low redshift and high redshift sample, after bifurcating the dataset at $z=1.4$. Our results for the best-fit values  for all the six regression relations can be found in Table~\ref{tab:parameters}. The marginalized credible intervals are shown in Fig.~\ref{fig_contour}, whereas the scatter plots for the  six regression relations are shown in Fig.~\ref{fig_xy}. Our conclusions are as follows:
\begin{itemize}
\item The slopes  and intercepts agree with the corresponding results with T21 for both the low and high redshift as well as the full sample.
\item Our intrinsic scatter for almost all the scaling relations are comparable to that found in T21. The only exception is the Amati relation where we see a much higher intrinsic scatter
compared to T21.
\item Although there is negligible redshift evolution in the scaling relations,  the high intrinsic  scatter implies that we cannot use these GRB observables for model-independent estimate of cosmological parameters.
\item We should note that another possibility for the large scatter could be that the quasar dataset  in ~\cite{Lusso20} cannot be used as a distance anchor~\cite{Ratra1,Ratra2,Ratra3} or the GRB dataset is not standardizable. One possibility would be to use the quasar  reverberation-mapped  observations discussed in ~\cite{Cao22} as distance anchors upto redshift of $z=3.4$.
\end{itemize}

In the spirit of open science, we have made available all our analysis codes and data publicly available at \url{https://github.com/saaarvesh/MTechThesis/tree/main/Final}. In a future work, we shall also update the analysis using the latest SWIFT data.

\section*{Acknowledgements}
This work builds upon the M.Tech thesis of Shreeprasad Bhat and we are grateful to him for sharing his codes. We are grateful to Bharat Ratra for comments on the manuscript. We also acknowledge the anonymous referees and editor  for useful feedback on the manuscript.

\bibliography{main}
\end{document}